\documentstyle[11pt,preprint,eqsecnum,aps,epsf]{revtex}
\preprint{RCNP-Th01006}

\newcommand{\be}{\begin{equation}}
\newcommand{\ee}{\end{equation}}
\newcommand{\bea}{\begin{eqnarray}}
\newcommand{\eea}{\end{eqnarray}}
\newcommand{\bvec}[1]{\mbox{\boldmath $#1$}}

\begin{document}

\draft
\title{Casimir scaling in a dual superconducting\\
scenario of confinement}

\author{Y. Koma$^{1,2}$\footnote{Email 
address: koma@rcnp.osaka-u.ac.jp}, 
E. -M. Ilgenfritz$^{1,3}$, H. Toki$^{1}$, and T. Suzuki$^{2}$}

\address{
$^{1}$ Research Center for Nuclear Physics (RCNP), Osaka University,\\ 
Mihogaoka 10-1, Ibaraki, Osaka 567-0047, Japan\\
\vspace{0.3cm}
$^{2}$ Institute for Theoretical Physics, Kanazawa University, 
Kanazawa 920-1192, Japan\\
\vspace{0.3cm}
$^{3}$ Institut f\"ur Theoretische Physik, 
Universit\"at T\"ubingen, D-72076 T\"ubingen, Germany}

\date{\today}

\maketitle

\begin{abstract}\baselineskip = 0.6cm
The string tensions of flux tubes associated with static 
charges in various SU(3) representations are studied within the 
dual Ginzburg-Landau (DGL) theory.
The ratios of the string tensions between higher and fundamental 
representations, $d_{D} \equiv \sigma_{D}/\sigma_{F}$, are 
found to depend only on the Ginzburg-Landau (GL) parameter, 
$\kappa = m_{\chi}/m_{B}$,
the mass ratio between monopoles $m_\chi$ and 
dual gauge bosons $m_B$.
In the case of the Bogomol'nyi limit ($\kappa=1$), analytical 
values of $d_{D}$ are easily obtained by adopting the manifestly 
Weyl invariant formulation of the DGL theory, which are provided 
simply by the number of color-electric Dirac strings inside the flux tube.
A numerical investigation of the ratio for various GL-parameter cases is 
also performed, which suggests that the Casimir scaling is obtained in the 
type-II parameter range within the interval $\kappa=5 \sim 9$ for various 
ratios $d_D$.\\
\end{abstract}


\baselineskip = 0.6cm

\section{Introduction}\label{sec:intro}

\par 
The observation of 
Casimir scaling is an important argument in any discussion 
of the virtues of different QCD vacuum models as far as the respective 
confinement mechanism is concerned.
Taken literally, the Casimir scaling suggests that the potential at 
intermediate distances between static charges in different
representation are proportional
to the eigenvalues $C^{(2)}(D)$ of the quadratic Casimir operator $T^a T^a$ 
in the respective $D$ dimensional representation, such that
$F_{D_1}(r)/F_{D_2}(r)=C^{(2)}(D_1)/C^{(2)}(D_2)$ at all distances.
This property is obvious only for the one-gluon exchange component of the
static force. Although there is no asymptotically linearly rising potential 
for the higher representations, at intermediate distances a string tension can
be defined which enters $F_{D}(r)$ as a constant part. 
The first lattice indications for the Casimir scaling appeared in the eighties
~\cite{Bernard,Ambjorn}. At that time this observation was a challenge
for the bag model~\cite{hansson}.
For example, the ratio of string tensions of adjoint to fundamental charges
in SU(3) gauge theory, respectively, would be 
$\sigma_{\mathrm adj}/\sigma_{\mathrm fund}=9/4=2.25$.

\par
Recently, as a contribution to the discussion of competing confinement 
mechanisms, Ref.~\cite{Deldar:1999vi} appeared where the string tensions of 
the fundamental and higher representations have been calculated in pure 
SU(3) lattice gauge theory, and the ratio 
was obtained nearly equal to 2, 
already rather close to 9/4.
In Ref.~\cite{Bali} Bali has studied the ratios of entire interaction 
potentials (including Coulomb and constant terms in addition to the 
linear term) also for quenched SU(3) gauge theory, and in the case 
of adjoint and fundamental 
charges the ratio turned out to be very close to 9/4.
All detailed (microscopic) mechanisms of confinement find it hard to 
explain the Casimir scaling, while it appears more natural from the 
point of view of the semi-phenomenological Stochastic Vacuum Model~\cite{SVM}. 
If the confinement mechanism is described by center vortices, 
approximate Casimir 
scaling for the potential can be achieved by introducing a finite thickness of 
the vortex, as demonstrated for the case of SU(2) lattice gauge 
theory~\cite{Faber:1998rp}, although the original center vortex picture 
gives a strictly vanishing potential for pairs of charges 
which transform trivially under $Z_N$ center of the gauge group.

\par 
For the dual superconductor scenario of 
confinement~\cite{Nambu:1974zg,Mandelstam:1974pi}, practically realized in 
the form of the dual Ginzburg-Landau (DGL) theory~\cite{Suzuki:1988yq}, one 
tends to believe that it would be difficult to accommodate Casimir scaling
in this framework. Indeed, in the Abelian projection 
scheme~\cite{tHooft:1981ht}
for SU(3) gluodynamics the long range forces are transmitted only
by ``diagonal gluons'' which couple to charges only via $T^3$ and $T^8$. 
This makes it hard to understand why the Casimir scaling should hold in  
Abelian projected gluodynamics. 
For example for the ratio between adjoint and fundamental 
forces one would naively expect the Abelian ratio equal to 3.  
As far as the derivation of the DGL theory is based on the Abelian projected 
gluodynamics, this seems to be unavoidable in the DGL theory, too.  
However, in a lattice investigation for SU(2) Abelian projected 
gluodynamics, 
Poulis~\cite{Poulis} has found the ratio between the string tensions of the
adjoint and fundamental representations, somewhere between the 
Abelian and Casimir scaling. 
This result is encouraging for the Abelian projected models to be able
to provide the Casimir scaling.
The case of SU(2) gluodynamics has been considered in 
Ref.~\cite{Chernodub:2000rg} in the context of an extended effective 
theory. It is discussed there that in the London limit Casimir scaling 
can be expected to hold. 
In this paper we examine straightforwardly the DGL theory for 
SU(3) gluodynamics with respect to the string tensions for 
various external charges without further modifications.

\par
Considering the DGL theory just at a phenomenological level, it might
be natural to restrict its application to 
mesonic~\cite{Suzuki:1988yq,Suganuma:1995ps,Koma:2000hw}, 
baryonic~\cite{Koma:2000hw,Kamizawa:1993hb}, 
glueball~\cite{Koma:1999sm} 
and perhaps to exotic states, and it would seem inappropriate to apply it 
to the so-called gluelump bound states made of infinitely heavy adjoint 
charges. However, because of the current interest in this issue, 
it is interesting to discuss how this kind of string would be 
represented within the DGL theory, and then to answer the question 
whether the Casimir scaling poses really a problem or not.

\par
In this paper we compute the string tensions of flux tubes which are 
originating from various dimensions of representation of color charge 
within the DGL theory. For this purpose we adopt the manifestly Weyl 
symmetric approach~\cite{Koma:2000hw,Koma:2000wn}, which will turn out 
to be very useful for classifying the flux tube in various representations.
Finally, based on these results as a function of the mass ratio between
dual gauge bosons and monopoles, we would like to discuss how 
good the Casimir scaling can be accomplished within the DGL theory.

\section{Weyl symmetric formulation of the DGL theory}

The DGL Lagrangian \cite{Suzuki:1988yq} is given by
\footnote{\baselineskip = 0.6cm
\baselineskip = 0.6cm
Throughout this paper, we use the following notations:
Latin indices $i$, $j$ express the labels 1,2,3, which are not
to be summed over unless explicitly stated.
Boldface letters, which appear later, denote three-vectors.}
\bea
{\cal L}_{\rm DGL} &=&
-\frac{1}{4} 
\left ( (\partial \wedge \vec{B})_{\mu\nu}
+ e \vec{\Sigma}_{\mu\nu} \right )^2 \nonumber\\
&&
  +\sum_{i=1}^3 
  \left [ \left | \left (\partial_{\mu}+ig\vec{\epsilon}
_{i}{\cdot}\vec{B}_{\mu} \right )\chi_{i} \right |^2
-\lambda \left ( \left |\chi_{i} \right |^2-v^2 \right )^2 \right ],
\label{eqn:DGL}
\eea
where $\vec{B}_{\mu}$ and $\chi_{i}$ denote the dual gauge field with 
two components $(B_{\mu}^3, B_{\mu}^8)$ and the complex scalar monopole 
field, respectively. 
The quark current $\vec{j}_{\mu} = \bar{q}\gamma_{\mu} \vec{H} q$, 
where $\vec{H}=(T_{3},T_{8})$, is represented by the boundary of a nonlocal 
string term $\vec{\Sigma}_{\mu\nu}$, which expresses the color-electric Dirac 
string singularity through the modified dual Bianchi identity
$\partial^{\nu} {}^{*\!}\vec{\Sigma}_{\mu\nu}=\vec{j}_{\mu}$.
Note that $(\partial \wedge \vec{B})_{\mu\nu} \equiv 
\partial_{\mu} \vec{B}_{\nu}-\partial_{\nu}\vec{B}_{\mu}$
satisfies $\partial^{\nu} {}^{*\!}(\partial \wedge \vec{B})_{\mu\nu}=0$.
Since the diagonal component of the matrix $\vec{H}$ 
gives the weight vector of the SU(3) algebra $\vec{w}_{j}$ 
($j=1,2,3$), where 
$\vec{w}_1= \left (1/2, \sqrt{3}/6 \right ), \vec{w}_2= \left
(-1/2, \sqrt{3}/6 \right ),  \vec{w}_3= \left (0, -1/\sqrt{3} \right )$, 
one can define the color-electric charges of the quarks as
$\vec{Q}_{j}^{(e)} \equiv e \vec{w}_{j}$.
Here, $j=1,2,3$ correspond to the color-electric charges, 
red ($R$), blue ($B$), and green ($G$).
Accordingly we can write the nonlocal term as
$e \vec{\Sigma}_{\mu\nu}= e \vec{w}_{j}\Sigma_{j\;\mu\nu}^{(e)}$.
On the other hand, 
the root vectors of the SU(3) algebra $\vec{\epsilon}_i$ are used 
to define the color-magnetic charges of the monopole field as 
$Q_{i}^{(m)} \equiv g \vec{\epsilon}_i$ ($i=1,2,3$), where
$\vec{\epsilon}_1=\left (-1/2,\sqrt{3}/2 \right ), 
  \vec{\epsilon}_2=\left(-1/2,-\sqrt{3}/2 \right )
, \vec{\epsilon}_3=\left (1,0 \right )$.
Both color-electric and color-magnetic charges satisfy the extended Dirac 
quantization condition 
$\vec{Q}_{i}^{(m)} \cdot \vec{Q}_{j}^{(e)} = 2\pi  m_{ij}$ ($eg=4\pi$).
Here $m_{ij}$ is an integer following the definition
\be
m_{ij} =2 \vec{\epsilon}_i \cdot \vec{w}_{j}
= \sum_{k=1}^{3} \epsilon_{ijk}=\{0,1,-1\},
\ee
where $\epsilon_{ijk}$ is the 3rd-rank antisymmetric tensor.
Typical mass scales in the DGL theory are 
the mass of the dual gauge field $m_B=\sqrt{3}gv$ and 
of the monopole field $m_\chi=2\sqrt{\lambda}v$.
Their ratio, the so-called Ginzburg-Landau (GL)
parameter $\kappa \equiv m_\chi/m_B$, characterizes the type of
dual ``superconductivity'' of the vacuum. Like in real superconductive 
materials, the properties of the vacuum might be very different 
depending on the actual value of $\kappa$.

\par
We make the Weyl symmetry of the DGL theory (\ref{eqn:DGL}) manifest, 
with the help of an extended dual gauge field \cite{Koma:2000hw,Koma:2000wn}, 
defined by 
\bea
B_{i\; \mu} \equiv g \vec{\epsilon}_{i} \cdot \vec{B}_{\mu}.
\quad\quad (i=1,2,3)
\label{eqn:b-weyl}
\eea
Here, a constraint $\sum_{i=1}^3 B_{i\; \mu} =0$ appears, since
$\sum_{i=1}^3 \vec{\epsilon}_{i} =0$. The DGL Lagrangian (\ref{eqn:DGL}) 
is now written as
\bea
{\cal L}_{\rm DGL} 
=
\sum_{i=1}^3 
\left [
- \frac{1}{4g_{m}^{2}} {{}^* \! F_{i\;\mu \nu}}^2
+
\left | \left (\partial_{\mu} + i B_{i\;\mu}\right )\chi_{i} \right |^2
- \lambda \left ( \left |\chi_{i} \right |^2-v^2 \right )^2 
\right ],
\label{eqn:dgl-weyl}
\eea
\bea
{}^* \! F_{i\;\mu \nu}\equiv  (\partial \wedge B_{i})_{\mu\nu}
 +2\pi \sum_{j=1}^{3} m_{ij} \Sigma_{j\;\mu\nu}^{(e)},
\eea
where the dual gauge coupling $g$ is scaled as
\be
g_{m} \equiv \sqrt{\frac{3}{2}} g.
\ee
The factor $2\pi$ in front of the Dirac string term is derived from 
the Dirac quantization condition.
Clearly, the expression (\ref{eqn:dgl-weyl}) is manifestly 
Weyl symmetric since all indices $i$ and $j$ are summed over.
Apparently the dual gauge symmetry is extended to 
[U(1)]$^{3}$, achieved by a set of transformation
\bea
&&
\chi_{i} \to  \chi_{i} e^{if_{i}},\quad
\chi^*_{i} \to \chi^*_{i} e^{-if_{i}},\nonumber\\
&&
B_{i\;\mu}^{\rm reg} \to
B_{i\;\mu}^{\rm reg} -\partial_{\mu}f_i, \quad\quad (i=1,2,3).
\label{eqn:gauge-sym-weyl}
\eea
However, the number of gauge degrees of freedom is not enlarged
because of the constraint $\sum_{i=1}^3 B_{i\;\mu} = 0$.

\par
In what follows we investigate the flux-tube solutions 
related to a separated quark and antiquark pair and related to 
analogous states (with higher representation charges) within the DGL theory.
In order to find such solutions it is useful to 
dispose the behavior of the dual gauge field, which can be 
achieved by the decomposition of the dual gauge field into two parts, 
the regular (no Dirac string) part and the singular (Dirac string) 
part \cite{Koma:2000wn},
\be
B_{i\;\mu} \equiv B_{i\;\mu}^{\rm reg} 
+ \sum_{j=1}^{3} m_{ij} B_{j\;\mu}^{\rm sing}
\quad (i=1,2,3),
\ee
where the singular part is determined so as to define
the color-electric charge density $C_{j\;\mu\nu}^{(e)}$ as
\be
 (\partial \wedge B_{j}^{\rm sing})_{\mu\nu}
+2 \pi \Sigma_{j\; \mu \nu}^{(e)}= 2\pi C_{j\; \mu \nu}^{(e)}, \quad
(j=1,2,3).
\label{eqn:su3-dg-singular-part}
\ee
The explicit form of $C_{j\; \mu \nu}^{(e)}$ is given by
\be
C_{j\; \mu\nu}^{(e)}(x) = \frac{1}{4\pi^{2}} \int d^{4}y \frac{1}{|x-y|^{2}}
{}^{*\!}(\partial \wedge j_{j}^{(e)}(y))_{\mu\nu},
\label{eqn:dah-Coulomb-term}
\ee
where $j_{j \; \mu}^{(e)}=\partial^{\nu} {}^{*\!}\Sigma_{j\;\mu\nu}^{(e)}$.
Note that if there is no quark source, we do not need to 
have a singular part $B_{j\;\mu}^{\rm sing}$.
Thus, the dual field strength tensor is rewritten as
\be
{}^{*\!}F_{i\;\mu\nu} = (\partial \wedge B_{i}^{\rm reg})_{\mu\nu} 
+2 \pi \sum_{j=1}^3 m_{ij} C_{j\;\mu\nu}^{(e)}.
\label{eqn:dah-dual-field-tensor-2}
\ee
In the static $q$-$\bar{q}$ system, $C_{j\;\mu \nu}^{(e)}$ turns out to be the 
Coulombic color-electric field originating from the color-electric charge.

\section{The string tension of the flux tube}

\par 
Let us now consider an idealized system, an infinitely long flux tube 
with cylindrical and translational symmetry. 
In this case, the terms related to $C_{j\;\mu \nu}^{(e)}$ can be neglected 
since they are relevant only for short separation of quark and antiquark.
One finds that the integration of square of $C_{j\;\mu \nu}^{(e)}$ 
gives the Coulomb energy including the self-energy of the 
color-electric charge. 
Correspondingly, now we only pay attention to the energy per length
(string tension) of the flux tube which has the terminating charges 
at infinity. In order to classify the types of the flux tube, 
we use a notation analogous to the $q$-$\bar{q}$ system.
The fields depend only on the radial coordinate $r$ as
\be
\phi_i = \phi_i(r),\quad
\bvec{B}_{i}^{\rm reg} = B_{i}^{\rm reg}(r)\bvec{e}_{\varphi} 
\equiv  
\frac{\tilde B_{i}^{\rm reg}(r)}{r}\bvec{e}_{\varphi},
\ee
where $\phi_i(r)$ is the modulus of 
the monopole field $\chi_i = \phi_{i} \exp (i\eta_{i})$, and
$\varphi$ denotes the azimuthal angle.
Note that the phase of the monopole field is now assumed to be
regular $[\partial_{\mu},\partial_{\nu}]\eta_{i}=0$, 
which is absorbed into the regular part of the dual gauge field by 
the replacement $B_{i}^{\rm reg} + \partial_{\mu} \eta_{i} \to 
B_{i}^{\rm reg}$.
If $[\partial_{\mu},\partial_{\nu}]\eta_{i}\ne 0$, this produces the
closed color-electric Dirac string singularity \cite{Koma:1999sm}.
Putting the quark at $z=-\infty$ and antiquark at $z=\infty$ (this
leads to the factor minus in $\bvec{B}_{i}^{\rm sing}$), the solution 
of (\ref{eqn:su3-dg-singular-part}) is easily found to be
\be
\bvec{B}_{i}^{\rm sing} 
=
-\frac{n_i^{(m)}}{r} \bvec{e}_{\varphi}.
\label{eqn:b-sing-inf}
\ee
Here $n_{i}^{(m)}$ is an integer corresponding to the number of 
color-electric Dirac strings in the dual gauge field within
the Weyl symmetric representation, which is expressed by the relation
\be
n_i^{(m)} \equiv \sum_{j=1}^{3}m_{ij}n_{j}^{(e)}.
\label{eqn:winding-number}
\ee
Here $n_j^{(e)}$ is the modulo $2\pi$ of $\Sigma_{j\;\mu\nu}^{(e)}$, 
the number of $j$-type color-electric charges attached to both 
ends of the flux tube. Various dimensions of the representation
of charges in SU(3) group and corresponding winding numbers are 
classified in Table.~\ref{tab:class-of-winding-number}.
For instance, the fundamental representation ($D={\bf 3}$) has three different 
charges $R$, $B$, and $G$.
These charges have the numbers $(n_1^{(e)},n_2^{(e)},n_3^{(e)})=(1,0,0)$,
$(0,1,0)$, and $(0,0,1)$, which are also written by using 
the expression (\ref{eqn:winding-number}) as
$(n_1^{(m)},n_2^{(m)},n_3^{(m)})=(0,-1,1)$, 
$(1,0,-1)$, and $(-1,1,0)$, respectively.
These rules hold similarly for the higher dimension of the representation.
However one should take into account the relation
$RB=\bar{G}$, $BG=\bar{R}$, $GR=\bar{B}$ and $RBG=0$ following
the definition of the fundamental color-electric charge 
with the weight vectors of the SU(3) algebra.
Making use of these, some of the charges classified in the 
higher dimension of the representation 
are reduced into that of the lower ones.

\par
Then, the field equations are the following
\bea
&&
  \frac{d^2 \tilde B^{\rm reg}_{i}}{dr^2} 
- \frac{1}{r}\frac{d \tilde B^{\rm reg}_{i}}{dr}
-2 g_{\rm m}^2 \left ( \tilde B^{\rm reg}_{i} - 
n_{i}^{(m)} \right ) 
\phi_i^2 = 0 ,
\label{eqn:f-eq-cyli-1}\\
&&\nonumber\\
&&
\frac{d^2\phi_i}{dr^2} +\frac{1}{r}\frac{d\phi_i}{dr}
-
\left ( 
\frac{ \tilde B^{\rm reg}_{i} - n_{i}^{(m)}}
{r} \right )^2 \phi_i 
- 2 \lambda \phi_i (\phi_i^2 - v^2) = 0.
\label{eqn:f-eq-cyli-2}
\eea
The string tension is the energy of the flux tube per unit length,
\bea
\sigma_D 
&=& 
2\pi 
\sum_{i=1}^3 
\int_0^{\infty}rdr
\Biggl [
\frac{1}{2 g_{\rm m}^2} \left ( \frac{1}{r}\frac{d \tilde B^{\rm reg}_{i}}
{dr} \right )^2
+
\left ( \frac{d \phi_i}{d r} \right)^2
\nonumber\\*
&&
+
\left ( 
\frac{ \tilde B^{\rm reg}_{i} -  n_{i}^{(m)}}{r} 
\right )^2 \! \phi_i^2
+
\lambda ( \phi_i^2 - v^2 )^2
\Biggr ],
\label{eqn:string-tension}
\eea
To make the energy of the system finite, we have to postulate 
the boundary conditions:
\bea
&&
\tilde B^{\rm reg}_{i} = 0, \quad \phi_i = 
\left  \{
\begin{array}{cc}
0 & \left (n_{i}^{(m)} \ne 0\right )\\
v & \left (n_{i}^{(m)} = 0 \right )
\end{array}
\right.
\qquad {\rm as}\qquad r \to 0, 
\nonumber\\
&&
\tilde B^{\rm reg}_{i} 
= n_{i}^{(m)},\quad \phi_i = v 
\qquad {\rm as}\qquad  r \to \infty.
\label{eqn:boundary-condition}
\eea

\par
For an analytical evaluation of the string tension,
it is useful to rewrite the expression (\ref{eqn:string-tension}) 
in the form \cite{Koma:2000wn},
\bea
\sigma_D
&=& 
2\pi v^2 \sum_{i=1}^{3}\left | n_i^{(m)} \right |
+
2\pi 
\sum_{i=1}^3 
\int_0^{\infty} \!\! r dr
\Biggl [
\frac{1}{2g_{\rm m}^2} 
\left ( \frac{1}{r}\frac{d \tilde B^{\rm reg}_{i}}{dr} 
\pm g_{\rm m}^2 ( \phi_i^2 - v^2 )
\right )^2 \nonumber\\
&&
+
\left ( \frac{d \phi_i}{d r} 
\pm
\left ( \tilde B^{\rm reg}_{i} -  n_i^{(m)} \right )
\frac{\phi_i}{r}
\right)^2 
+
\frac{1}{2}\left (2 \lambda - g_{\rm m}^2 \right ) ( \phi_i^2 - v^2 )^2
\Biggr ].
\label{eqn:string-tension-2}
\eea
{}From this expression we find that in the Bogomol'nyi limit 
\cite{Koma:2000wn,Chernodub:1999xi},
\be
g_{\rm m}^2 = 2\lambda, \quad {\rm or}\quad 3 g^2 = 4 \lambda,
\label{eqn:bogomolnyi-limit}
\ee
which corresponds to $\kappa=m_\chi/m_B=1$, 
a considerable simplification occurs.
The vacuum is separated into two types at this Bogomol'nyi point,
type-I ($\kappa<1$) and type-II ($\kappa>1$) in analogy to the 
superconducting material.
In the Bogomol'nyi limit one gets the saturated string tension, 
\be
\sigma_D
= 
2 \pi v^2 \sum_{i=1}^3 \left | n_{i}^{(m)} \right |.
\label{eqn:st-exact}
\ee
One finds that the string tension for the fundamental representation
becomes $\sigma_3 = 4\pi v^2$ since 
$\sum_{i=1}^3 \left | n_{i}^{(m)} \right |=2$.
In this special case, the profiles of the dual gauge field and the 
monopole field are determined by the first-order differential equations,
\bea
&&
\frac{1}{r} \frac{d \tilde B^{\rm reg}_{i} }{dr}
\pm
g_{\rm m}^2 (\phi_i^2 - v ^2) =0,
\label{eqn:bogomol-eq-1}\\
&&
\frac{d\phi_i}{dr} 
\pm 
\left ( \tilde B^{\rm reg}_{i}  - n_{i}^{(m)} \right )
\frac{\phi_i}{r}=0.
\label{eqn:bogomol-eq-2}
\eea
These field equations reproduce the second-order 
differential equations (\ref{eqn:f-eq-cyli-1}) and
(\ref{eqn:f-eq-cyli-2}) when the relation 
(\ref{eqn:bogomolnyi-limit}) is fulfilled.
Note that the procedure to find the Bogomol'nyi limit 
is an extension of the method used for 
the U(1) Abelian Higgs model~\cite{Bogomolny:1976de,deVega:1976mi}
to the U(1)$\times$U(1) DGL theory corresponding to SU(3) 
gluodynamics in Abelian projection.

\par
\begin{figure}[hbt]
\centerline{\epsfxsize = 8cm\epsfbox{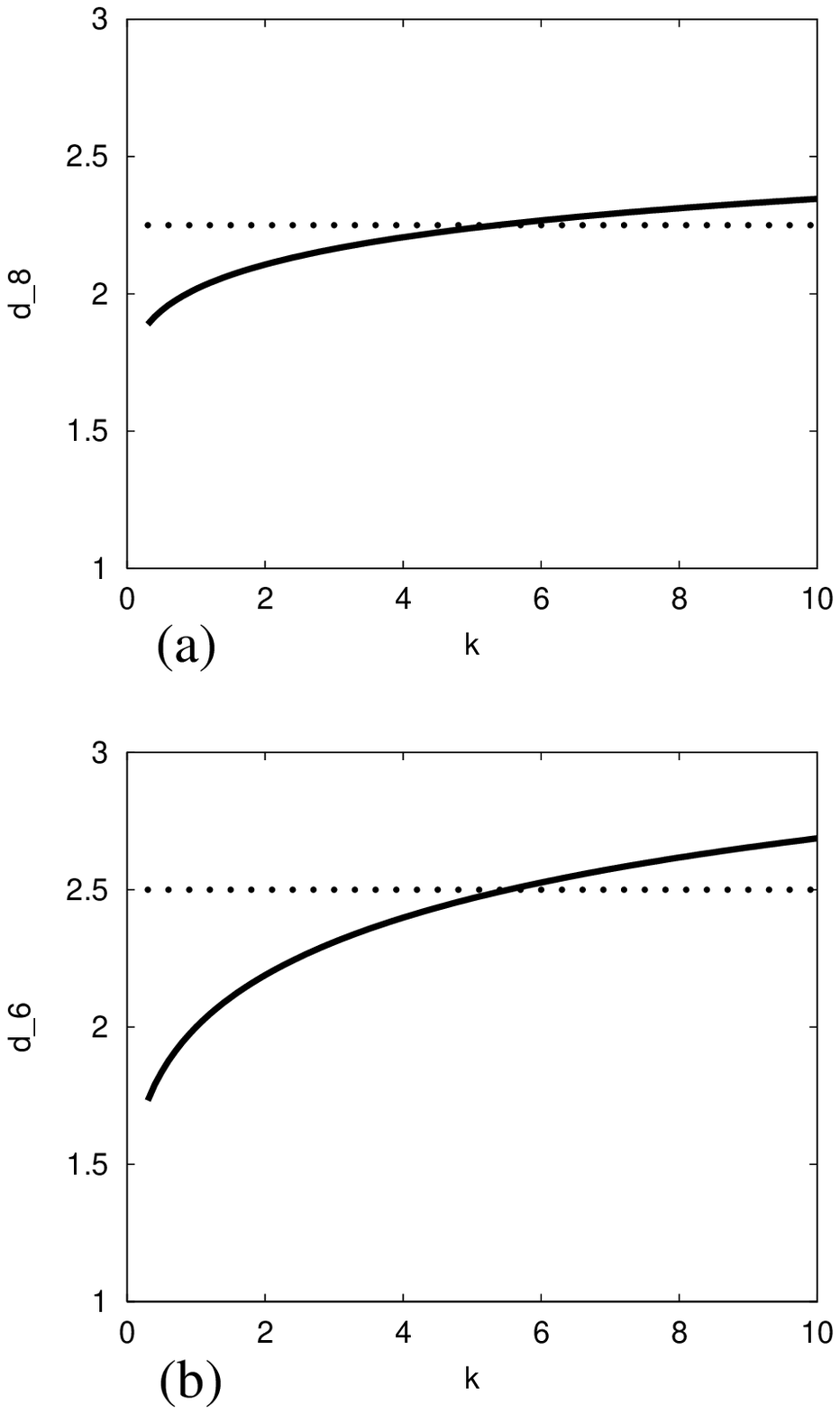}}
\caption{\small 
The ratio of the string tension 
(a) for the octet representation $d_8 = \sigma_8/\sigma_3$ 
(the dotted line marks the ratio of quadratic Casimir charges 
$C^{(2)}({\bf 8})/C^{(2)}({\bf 3})=2.25$)
and (b) for the sextet representation $d_6 = \sigma_6/\sigma_3$
($C^{(2)}({\bf 6})/C^{(2)}({\bf 3})=2.5$).
The weight factor is $p+q=2$.
Casimir scaling for the values of $d_8$ and $d_6$ 
is observed at $\kappa \approx 5$.}
\label{fig:d8-d6}
\end{figure}
\begin{figure}[hbt]
\centerline{\epsfxsize = 8cm\epsfbox{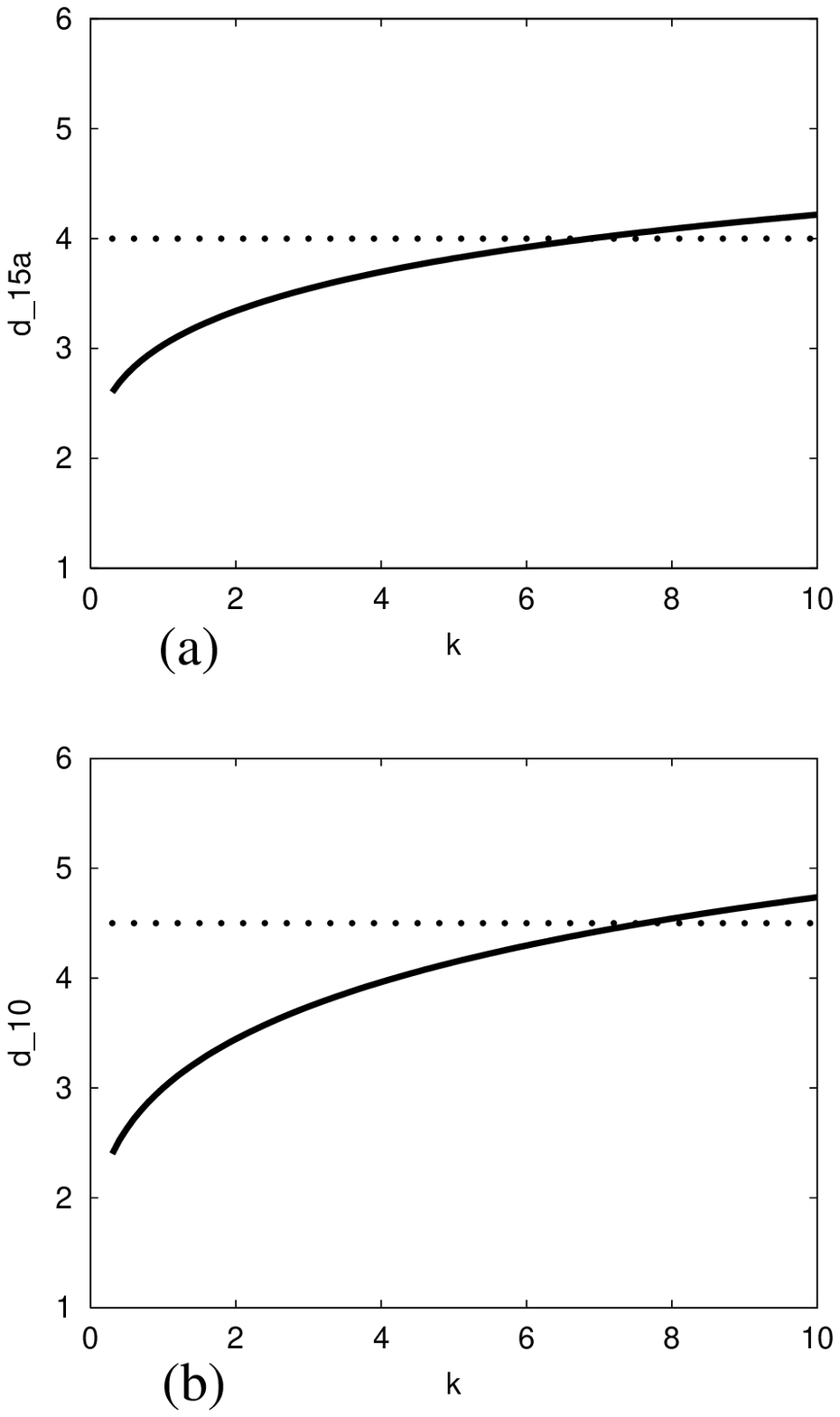}}
\caption{\small 
Similar ratios as in Fig.~\ref{fig:d8-d6}, 
(a) $d_{15a} = \sigma_{15a}/\sigma_3$ 
($C^{(2)}({\bf 15a})/C^{(2)}({\bf 3})=4$)
and (b) $d_{10} = \sigma_{10}/\sigma_3$
($C^{(2)}({\bf 10})/C^{(2)}({\bf 3})=4.5$).
The weight factor is $p+q=3$.
Casimir scaling for the values of $d_{15a}$ and $d_{10}$ 
is observed at $\kappa \approx 7$.}
\label{fig:d15a-d10}
\end{figure}
\begin{figure}[hbt]
\centerline{\epsfxsize = 8cm\epsfbox{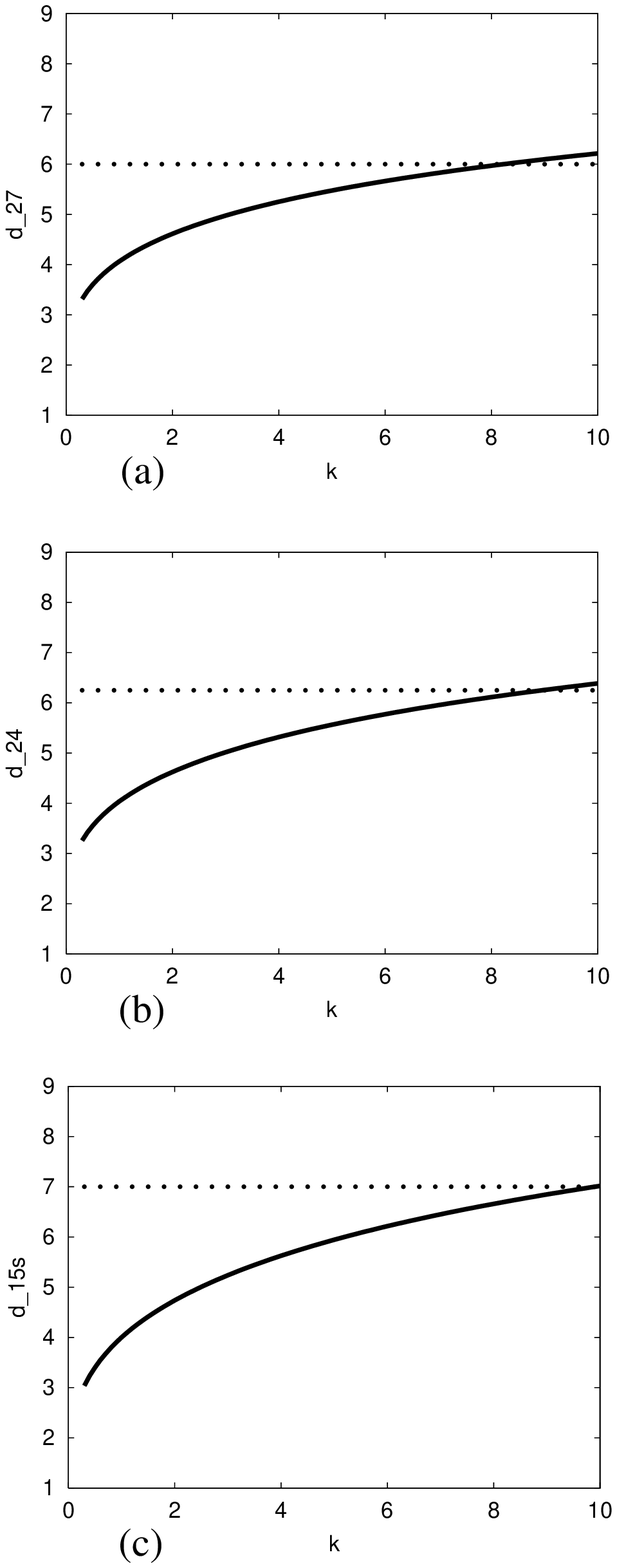}}
\caption{\small 
Similar ratios as in Fig.~\ref{fig:d8-d6}, 
(a) $d_{27} = \sigma_{27}/\sigma_3$ 
($C^{(2)}({\bf 27})/C^{(2)}({\bf 3})=6$), 
(b) $d_{24} = \sigma_{24}/\sigma_3$
($C^{(2)}({\bf 24})/C^{(2)}({\bf 3})=6.25$), 
and (c)  $d_{15s} = \sigma_{15s}/\sigma_3$
($C^{(2)}({\bf 15s})/C^{(2)}({\bf 3})=7$). 
The weight factor is $p+q=4$.
Casimir scaling for the values of $d_{27}$, $d_{24}$, and $d_{15s}$ 
is observed at $\kappa \approx 9$.}
\label{fig:d27-d24-d15s}
\end{figure}

Finally, let us compute the ratio of the string tension between 
the higher and fundamental representations.
In the Bogomol'nyi limit, this can be done easily by using the expression
(\ref{eqn:st-exact}).
For the ratio between $D={\bf 8}$ and $D={\bf 3}$, we get
$d_8 \equiv \sigma_8/\sigma_3 = (2\pi v^2\times 4)/ (2\pi v^2\times 2)=2$.
In general, one can recognize a simple rule
\be
d_D \equiv \frac{\sigma_D}{\sigma_3} = p+q.
\label{eqn:string-tension-ratio}
\ee
Here $p+q$ is the sum of weight factors in the SU(3) representation, 
which physically corresponds to the number of 
the color-electric Dirac string inside the flux tube
in the framework of the DGL theory.
In the type-I ($\kappa<1$) or type-II ($\kappa>1$) 
parameter range, we have to calculate
the expression (\ref{eqn:string-tension}) by solving the 
field equations (\ref{eqn:f-eq-cyli-1}) and (\ref{eqn:f-eq-cyli-2})
numerically.
The corresponding numerical results are shown 
in Figs.~\ref{fig:d8-d6}$-$\ref{fig:d27-d24-d15s}.
In Fig.~\ref{fig:d8-d6} we show the values of $d_8$ and $d_6$
corresponding to $p+q=2$ as a function of the GL parameter.
Similarly, in Figs.~\ref{fig:d15a-d10} and~\ref{fig:d27-d24-d15s}, 
the ratios $d_D$ classified in the representation $p+q=3$ and $p+q=4$
are shown, respectively. 
The dotted line denotes the ratio of the quadratic Casimir charges:
$C^{(2)}(D)/C^{(2)}(F={\bf 3})=2.25$, $2.5$, $4$, $4.5$, $6$, $6.25$, $7$
for the dimensions $D={\bf 8}$, ${\bf 6}$, ${\bf 15a}$, ${\bf 10}$,
 ${\bf 27}$, ${\bf 24}$, ${\bf 15s}$, respectively.
Of course, at $\kappa=1$ the analytical results 
(\ref{eqn:string-tension-ratio}) are realized, and
all ratios increase monotonously with the GL parameter $\kappa$.
It is interesting to note that the Casimir scaling 
is accomplished only in the type-II region, for $\kappa=5 \sim 9$, depending
on the Casimir ratio for a particular representation.
It seems that the DGL theory in its present shape can not describe the
Casimir scaling for all representations simultaneously, with a unique
value of $\kappa$. Some additional terms will be necessary to slightly
modify the influence of $\kappa$ on the energy of the flux tube solutions
corresponding to various external charges.

\section{Summary and discussion}

We have studied the string tension of flux tubes associated 
with static charges in higher SU(3) representations within the 
dual Ginzburg-Landau theory in a manifestly Weyl symmetric approach.
We have found that the ratio of the string tension between 
higher and fundamental representations, 
$d_{D} \equiv \sigma_{D}/\sigma_{F}$,  depends only 
on the ratio  between the monopole mass $m_\chi$ and 
the mass of the dual gauge boson $m_B$, 
the Ginzburg-Landau parameter $\kappa = m_{\chi}/m_{B}$.
We have pointed out that the ratios $d_{D}$ have a simple form in the case 
of the Bogomol'nyi limit in terms of the number of color-electric Dirac 
strings inside the flux tube. We have numerically determined the ratios 
in the type-I ($\kappa<1$) and type-II ($\kappa>1$) parameter ranges
and found them monotonically rising with $\kappa$.
In principle, such deviation of the ratio from the number of
color-electric Dirac string can be understood 
as the effect of the flux-tube interaction.
In the type-II parameter range, in the interval $\kappa=5 \sim 9$, 
the Casimir scaling is approximately reproduced for all 
representations $D$ studied in the present paper.
This shows that it was premature to say that the
the dual superconducting scenario of confinement 
is in plain contradiction with the Casimir scaling. 

\section*{Acknowledgment}
The authors acknowledge fruitful discussions with M. Takayama.
E.-M.I. is grateful for the support by the 
Ministry of Education, Culture and Science of Japan (Monbu-Kagaku-sho).
T.S. acknowledges the financial support from JSPS Grant-in Aid for
Scientific Research (B) No. 10440073 and No. 11695029.

\newpage
\begin{table}
\begin{center}
\begin{minipage}{12cm}
\caption{\small 
Classification of the color-electric charges and 
the winding numbers of flux tubes
for various dimensions of the representations of SU(3) group.
$D$ denotes the dimension of representation, ($p,q$) the weight
factors. $Q$ is the color-electric charge of $Q$-$\bar{Q}$ flux-tube 
system, where the number of the color-electric Dirac strings 
attached to the charge $Q$ is $n_j^{(e)}$. 
The number of strings in the dual gauge field within
the Weyl symmetric representation is given by 
$n_i^{(m)}=\sum_{j=1}^3 m_{ij}n_j^{(e)}$, where 
$m_{ij}=2\vec{\epsilon}_i\cdot \vec{w}_j$.}
\label{tab:class-of-winding-number}
\vspace{0.3cm}
\end{minipage}
\begin{tabular}{ccclcrrrrrrrr}
 $D$ & $(p,q)$ & $p+q$ 
 &\multicolumn{1}{c}{$Q$}
&&\multicolumn{1}{c}{$n_{1}^{(e)}$}
 &\multicolumn{1}{c}{$n_{2}^{(e)}$}
 &\multicolumn{1}{r}{$n_{3}^{(e)}$}
&&\multicolumn{1}{r}{$n_{1}^{(m)}$}
 &\multicolumn{1}{r}{$n_{2}^{(m)}$}
 &\multicolumn{1}{r}{$n_{3}^{(m)}$}\\
\hline
{\bf 3}
&(1,0)&1& $R$ && $1$ & $0$ & $0$ &&  $0$ & $-1$ &  $1$\\
&     & & $B$ && $0$ & $1$ & $0$ &&  $1$ &  $0$ & $-1$\\
&     & & $G$ && $0$ & $0$ & $1$ && $-1$ &  $1$ &  $0$\\
\hline
{\bf 8}
&(1,1)&2&$R\bar{B}$ &&  $1$ & $-1$ &  $0$ && $-1$ & $-1$ & $2$\\
&     &  & $B\bar{G}$ &&  $0$ &  $1$ & $-1$ &&  $2$ & $-1$ & $-1$\\
&     &  & $G\bar{R}$ && $-1$ &  $0$ &  $1$ && $-1$ &  $2$ & $-1$\\
\hline
{\bf 6}
&(2,0)&2& $RR$ && $2$ & $0$ & $0$ &&  $0$ & $-2$ &  $2$\\
&     & & $BB$ && $0$ & $2$ & $0$ &&  $2$ &  $0$ & $-2$\\
&     & & $GG$ && $0$ & $0$ & $2$ && $-2$ &  $2$ &  $0$ \\
\hline
{\bf 15a}
&(2,1)&3& $RR\bar{B}$ &&  $2$ & $-1$ &  $0$ && $-1$ & $-2$ &  $3$\\ 
&     & & $BB\bar{G}$ &&  $0$ &  $2$ & $-1$ &&  $3$ & $-1$ & $-2$\\ 
&     & & $GG\bar{R}$ && $-1$ &  $0$ &  $2$ && $-2$ &  $3$ & $-1$\\
&     & & $RR\bar{G}$ &&  $2$ &  $0$ & $-1$ &&  $1$ & $-3$ &  $2$\\
&     & & $BB\bar{R}$ && $-1$ &  $2$ &  $0$ &&  $2$ &  $1$ & $-3$\\
&     & & $GG\bar{B}$ &&  $0$ & $-1$ &  $2$ && $-3$ &  $2$ &  $1$\\
\hline
{\bf 10}
&(3,0)&3& $RRR$ && $3$ & $0$ & $0$ &&  $0$ & $-3$ &  $3$\\
&     & & $BBB$ && $0$ & $3$ & $0$ &&  $3$ &  $0$ & $-3$\\
&     & & $GGG$ && $0$ & $0$ & $3$ && $-3$ &  $3$ &  $0$\\
\hline
{\bf 27}
&(2,2)&4& $RR\bar{B}\bar{B}$ &&  $2$ & $-2$ &  $0$ && $-2$ & $-2$ &  $4$\\ 
&     & & $BB\bar{G}\bar{G}$ &&  $0$ &  $2$ & $-2$ &&  $4$ & $-2$ & $-2$\\
&     & & $GG\bar{R}\bar{R}$ && $-2$ &  $0$ &  $2$ && $-2$ &  $4$ & $-2$\\
\hline
{\bf 24}
&(3,1)&4& $RRR\bar{B}$ &&  $3$ & $-1$ &  $0$ && $-1$ & $-3$ &  $4$\\ 
&     & & $BBB\bar{G}$ &&  $0$ &  $3$ & $-1$ &&  $4$ & $-1$ & $-3$\\
&     & & $GGG\bar{R}$ && $-1$ &  $0$ &  $3$ && $-3$ &  $4$ & $-1$\\
&     & & $RRR\bar{G}$ &&  $3$ &  $0$ & $-1$ &&  $1$ & $-4$ &  $3$\\
&     & & $BBB\bar{R}$ && $-1$ &  $3$ &  $0$ &&  $3$ &  $1$ & $-4$\\
&     & & $GGG\bar{B}$ &&  $0$ & $-1$ &  $3$ && $-4$ &  $3$ &  $1$\\
\hline
{\bf 15s}
&(4,0)&4& $RRRR$ && $4$ & $0$ & $0$ &&  $0$ & $-4$ &  $4$\\
&     & & $BBBB$ && $0$ & $4$ & $0$ &&  $4$ &  $0$ & $-4$\\
&     & & $GGGG$ && $0$ & $0$ & $4$ && $-4$ &  $4$ &  $0$\\
\end{tabular}
\end{center}
\end{table}

\newpage

\begin{thebibliography}{10}

\bibitem{Bernard}
C.~Bernard, Nucl. Phys. {\bf B219}, 341 (1983).

\bibitem{Ambjorn}
J.~Ambjorn, P. Olesen, and C. Peterson, Nucl. Phys. {\bf B240}, 189 (1984).

\bibitem{hansson} 
T.~H.~Hansson, Phys. Lett. {\bf B166}, 343 (1986).

\bibitem{Deldar:1999vi}
S. Deldar, Phys. Rev. {\bf D62},  034509 (2000).
 
\bibitem{Bali}
G.~S.~Bali, Phys. Rev. {\bf D62}, 114503 (2000).

\bibitem{SVM}
V.~I.~Shevchenko and Yu.~A.~Simonov, Phys. Rev. Lett. {\bf 85}, 1811 (2000).

\bibitem{Faber:1998rp}
M. Faber and J. Greensite, and S. Olejnik, 
Phys. Rev. {\bf D57}, 2603 (1998).


\bibitem{Nambu:1974zg}
Y. Nambu, Phys. Rev. {\bf D10},  4262  (1974).

\bibitem{Mandelstam:1974pi}
S. Mandelstam, Phys. Rept. {\bf 23C},  245  (1976).

\bibitem{Suzuki:1988yq}
T. Suzuki, Prog. Theor. Phys. {\bf 80},  929  (1988);\\
S. Maedan and T. Suzuki, {\it ibid}. {\bf 81},  229  (1989).


\bibitem{tHooft:1981ht}
G. 't~Hooft, Nucl. Phys. {\bf B190},  455  (1981).
 

\bibitem{Poulis}
G.I. Poulis, Phys. Rev. {\bf D54},  6974 (1996).

\bibitem{Chernodub:2000rg}
M.~N.~Chernodub, F.~V.~Gubarev, M.~I.~Polikarpov, 
and V.~I.~Zakharov, hep-th/0010265.


\bibitem{Suganuma:1995ps}
H.~Suganuma, S.~Sasaki, and H. Toki, Nucl. Phys. {\bf B435}, 207 (1995).


\bibitem{Koma:2000hw}
Y. Koma, E.~M. Ilgenfritz, T. Suzuki, and H. Toki, 
preprint, hep-ph/0011165, to appear in PRD (2001).

\bibitem{Kamizawa:1993hb}
S. Kamizawa, Y. Matsubara, H. Shiba, and T. Suzuki, Nucl. Phys. {\bf B389},
  563  (1993).

\bibitem{Koma:1999sm}
Y. Koma, H.~Suganuma, and H. Toki, Phys. Rev. {\bf D60},  074024  (1999).

\bibitem{Koma:2000wn}
Y. Koma and H. Toki, Phys. Rev. {\bf D62},  054027  (2000).

\bibitem{Chernodub:1999xi}
M.~N. Chernodub, Phys. Lett. {\bf B474},  73  (2000).

\bibitem{Bogomolny:1976de}
E.~B. Bogomol'nyi, Sov. J. Nucl. Phys. {\bf 24},  449  (1976).

\bibitem{deVega:1976mi}
H.~J. de~Vega and F.~A. Schaposnik, Phys. Rev. {\bf D14},  1100  (1976).


\end{thebibliography}
\end{document}